# Compiling, Validating, and Using Human Reference Atlas Cell Type Annotation Crosswalks

Nicole Vasilevsky[1,2*], Aleix Puig-Barbe[3*], Andreas Bueckle[1], Ellen M. Quardokus[1], Yongxin Kong[1], Divya Prasanth Paraman[1], Jie Zheng[4], Yongqun He[4], Yashvardhan Jain[1], Bruce W. Herr II[1], Katy Börner[1*]

[1] Department of Intelligent Systems Engineering, Luddy School of Informatics, Computing, and Engineering, Indiana University, Bloomington, IN, USA
[2] Rose City Data Science, Portland, OR, USA
[3] European Bioinformatics Institute (EMBL-EBI), Wellcome Genome Campus, Hinxton, Cambridge CB10 1SD, UK
[4] University of Michigan Medical School, Ann Arbor, MI, USA

* Corresponding authors
nvasilevsky@c-path.org
aleix@ebi.ac.uk
katy@iu.edu

# Abstract

The Human Reference Atlas (HRA) v2.5 includes 4,807 anatomical structures, 1,347 cell types, and 2,215 biomarkers (genes, proteins, lipids); 81 organs with 1,392 anatomical structures modeled in three dimensions (3D). Mapping new data into the HRA requires 3D spatial registration and/or cell type annotation (CTann). Many CTann tools exist but few use cell type names linked to Cell Ontology (CL)—making comparisons across tools, tissues, and organs difficult. The HRA supports five transcriptomics CTann tools (Azimuth, CellTypist, popV, FR-Match, Pan-Human Azimuth), four sc-proteomics CTann tools (DeepCell Types, DeepCell Types-HuBMAP, Robust Image-Based Cell Annotator [RIBCA], and SpaTial cELl LeARning [STELLAR]), plus a Cell Distance Explorer (CDE) Spatial Omics collection covering previously published annotations across 12 tissue types with 47,349,496 cells. Assigned cell types differ in number, resolution, and coverage. This paper details how 10 crosswalks were constructed, validated, and used to compile a 'Supertree' that can be used to compare CTann across organs and technologies in support of reference atlas constitution and usage. All data and code can be accessed at https://cns-iu.github.io/hra-ctann-supporting-information.

# Background & Summary

The Human Reference Atlas (HRA)[1] is being developed by the NIH-funded Human BioMolecular Atlas Program (HuBMAP)[2,3], Cellular Senescence Network[4] (SenNet) Data Portal[5], Kidney Precision Medicine Project[6] (KPMP), Canadian Institute for Advanced Research (CIFAR) MacMillan Multiscale Human, and serves 25+ consortia. There are two main methods to map new data into the HRA so it can be explored in the spatial-semantic context of this reference and compared with 10,000+ other datasets: (1) registering human tissue samples spatially into the 3D male and female reference bodies using the Registration User Interface (RUI)[7] and/or (2) registering data semantically using HRA-supported cell type annotation (CTann) tools, crosswalked to Cell Ontology (CL)[8,9]. This paper focuses on the latter mapping; it details the compilation and validation of 10 crosswalks for single-cell (sc-)transcriptomics and sc-spatial omics data in support of HRA construction and usage. Formal definitions of CTann and related terminology are provided in the HRA Glossary[10].

As of August 2026, there is not a standardized approach to annotate cells in sc-transcriptomics and sc-spatial omics data. Different organ experts and CTann tool developers use different cell type names (e.g., 'T' or 'T-cell') at at different levels of granularity (e.g., 'EC aerocyte capillary' or 'alveolar capillary type 2 endothelial cell') to describe the same or closely related cell types. Relatively few expert-curated datasets and annotation tools use standardized CL labels and identifiers; one example is popV[11]. Even if terminology is aligned to CL, different assay types and the panel size used, for example, GeoMx imaging using a 3-plex antibody panel[12,13] versus PhenoCycler imaging using a 54-plex antibody panel[14], make it difficult to compute and compare cell type annotations (and cell type populations) across studies, as the technologies provide different amounts of information from which cell identity can be inferred. For example, when compared with human annotations, Pan-Human Azimuth annotations in lung show greater difficulty distinguishing certain cell types, including goblet cells, capillary and venous endothelial cells, fibroblast subtypes, pericytes and vascular smooth muscle cells, and specific immune cell subtypes. In contrast, cell types such as alveolar type 1 (AT1), alveolar type 2 (AT2), and mast cells are annotated more consistently. Major variation in how cells are annotated exists for rare cell types.

**For single-cell transcriptomics data**, diverse CTann tools exist to annotate data in the HuBMAP, SenNet, Genotype-Tissue Expression (GTEx) sc-portal[15], Chan Zuckerberg Initiative CELLxGENE (CxG) Portal[16] , Deeply Integrated Human Single-Cell Omics (DISCO)[17], Human Cell Atlas (HCA)[18,19], Human Tumor Atlas Network (HTAN) portals[20,21]. Prior work has examined the coverage and agreement between cell type annotation (CTann) tools. Hu et al.[22] benchmarked four CTann tools—Azimuth[23], CellTypist[24,25], scArches[26], and FR-Match[27]—using two lung atlas datasets: the filtered core subset of 584,944 healthy cells of 61 types from the Human Lung Cell Atlas (HLCA) dataset (2.4 million cells across 486 donors total)[28,29] and the LungMAP single-cell reference (CellRef) with 347,970 cells of 48 types from 104 donors[30]. The four CTann tools agree on annotations for 76.7% of cells when performing cell-based matching of the CellRef dataset as a query to the HLCA reference. As described by Hemberg et al.[31], there is no single unified way to perform preprocessing on sc-transcriptomics datasets for cell atlas creation, and batch effect mitigation[32–34] as well as label transfer[35–37] remain challenges in sc atlases that are actively being addressed. An overview of CTann tools and datasets used in the HRA is presented in **Table 1**.

**Table 1: CTann tools used in HuBMAP, SenNet, GTEx, CELLxGENE, DISCO, HCA, HTAN, and HRA to annotate cells in sc-transcriptomics and spatial proteomics data**

| Name of CTann Tool | Organs (# Cell IDs) | Code |
|---|---|---|
| **SC-Transcriptomics** | | |
| Azimuth | adipose tissue (26), adrenal gland (12), blood (41), bone marrow (43), brain (8), eye (16), heart (33), intestine (12), kidney (64), liver (32), lung (85), palatine tonsil (66), pancreas (22), placenta (12), primary motor cortex (19), skeletal muscle tissue (11), spleen (9), stomach (16), thymus (5), white adipose tissue (26) | https://azimuth.hubmapconsortium.org |
| CellTypist | blood (22), bone marrow (38), heart (50), hippocampal formation (22), intestine (39), islet of Langerhans (4), kidney (35), large intestine (87), liver (35), lung (73), lymph node (27), pancreas (9), skeletal muscle tissue (19), skin of body (23), small intestine (87), spleen (29) | https://www.celltypist.org |
| FR-Match | lung (53) | https://github.com/NLM-DIR/FRmatch |
| Pan-Human Azimuth | body proper (241) | https://satijalab.org/pan_human_azimuth |
| popV | adipose tissue (12), blood (20), blood vasculature (14), bone marrow (15), eye (30), heart (5), large intestine (18), liver (12), lung (40), lymph node (23), male reproductive system (13), mammary gland (14), mesenteric lymph node (25), pancreas (14), prostate gland (13), respiratory system (36), skin of body (24), small intestine (17), spleen (25), thymus (23), tongue (11), trachea (18), urinary bladder (15), uterus (13) | https://github.com/YosefLab/popV |
| **SC-Spatial Proteomics & SC-Spatial Omics** | | |
| CDE Spatial Omics+ | bone marrow (42), decidua (20), esophagus (18), intestine (31), lung (26), lymph node (30), oral cavity (34), pancreas (5), skin of body (17), spleen (14), tonsil (47) | https://github.com/cns-iu/hra-cell-distance-analysis |
| DeepCell Types | body proper (29) | https://github.com/vanvalenlab/deepcell-types |
| DeepCell Types-HuBMAP | body proper (40) | https://github.com/hubmapconsortium/deepcelltypes |
| RIBCA | body proper (15) | https://github.com/sun-huangqingbo/multiplexed-image-annotator |
| STELLAR | intestine (21 | https://snap.stanford.edu/stellar |

+Contains one sc-spatial omics Xenium dataset with sc-spatial transcriptomics and sc-proteomics data.

**For single-cell spatial proteomics and spatial omics data**, diverse cell segmentation algorithms[38–40] exist; segmentation results impact cell annotation but they are not the focus of this paper. HuBMAP uses STELLAR[41], DeepCell Types[42], and RIBCA[43] for annotating cell types in sc-spatial proteomics datasets. HRA v2.5 uses these three CTann tools plus annotations from the CDE Spatial Omics metastudy[3] (see **Table 1** and **Methods**).

**HRA Cell type Populations** (HRApop)[44] v1.1 uses five CTann tools (Azimuth, CellTypist, popV, FR-Match, and/or Pan-Human Azimuth[45]) to annotate 634 cell type labels (375 CL IDs) for 22 organs in sc-transcriptomics data; manual annotations are used to assign 105 cell type labels (93 CL IDs) for 9 organs in sc-proteomics data; resulting in a total of 724 unique cell type labels (395 CL IDs) for 103 anatomical structures (147 if male and female are counted separately) in 23 organs (38 if male and female are counted separately). HRApop v1.1 was published on Zenodo[46] and the HRA KG[47] on June 30, 2026. Note that some CL IDs have multiple matching CL labels; for example, CL ID http://purl.obolibrary.org/obo/CL_0000037 has the main label 'hematopoietic stem cell', the exact synonyms 'blood forming stem cell' and 'hemopoietic stem cell', and the related synonyms 'colony forming unit hematopoietic' and 'HSC'. All these labels share exactly one CL ID.

# Methods

To harmonize cell type annotations across datasets, CTann crosswalks, a collection of CSV files linking CTann-generated cell type labels to CL or Provisional Cell Ontology (PCL)[48,49] IDs, are used. These crosswalks enable alignment of CTann tool generated and human expert provided cell type annotations to organ-specific ASCT+B tables[50] and other digital objects in the HRA.

Subsequently, we detail sc-transcriptomics and sc-proteomics data, CTann tools and crosswalks for both data types, discuss inter-rater variability, and present a comparison of CT typologies per anatomical structure (AS) and organ in the CTann Supertree. The CTann Supertree provides a common hierarchical framework for organizing and comparing CT typologies across anatomical structures, organs, and annotation technologies.

We distinguish among three key concepts: CL terms, labels, and annotations. A CL term is an official Cell Ontology entry that corresponds to a specific cell type that is defined by a persistent CL ID and its official metadata (also referred to as a class by ontology developers). Key metadata includes a primary term label, a textual (human-readable) definition, and in some cases, a computational definition and synonyms. In contrast, a label is a human-readable name used by datasets or CTann tools; these may differ from the official CL term due to synonyms, abbreviations, or tool-specific formatting. Finally, a cell type annotation is the actual assignment of a label or CL term to specific cells based on experimental data.

## SC-Transcriptomics and SC-Proteomics/SC-Spatial Omics Data

The HRA v2.5 covers 34 ASCT+B tables[50] with 4,807 anatomical structures linked to anatomical ontologies Uberon Anatomy Ontology (Uberon)[51] or Foundational Model of Anatomy (FMA)[52], 1,347 cell types linked to CL, and 2,215 biomarkers linked to HGNC.

HRApop v1.1 sc-transcriptomics data covers 11,832,850 cells from 643 single-cell datasets. HRA spatial omics (mostly proteomics and some transcriptomics) data covers 47,349,496 cells from 399 spatially resolved omics datasets from 12 organs from 14 studies.

Sc-transcriptomics data used Azimuth, CellTypist, popV, FR-Match, and Pan-Human Azimuth for cell type annotation. Cell type labels for sc-proteomics data were procured by the data providers, then crosswalked. Some cell type annotation tools for sc-transcriptomics data provide cell types at different levels; Azimuth, for example, features two annotation levels for the heart and three for the kidney. For sc-proteomics data, a manually curated crosswalk[53] from original cell types to three higher-level aggregations of cell types was used.

While both the sc-transcriptomics and sc-proteomics datasets mapped to the HRA contain cell types at multiple levels of aggregation, their underlying cell type hierarchies were not harmonized until the supertree was developed.

## Selection of CTann Tools

There are 50+ cell type annotation tools for sc-transcriptomics data and 30+ for sc-proteomics and multi-omics data. Using different tools on the same dataset results in different cell type annotations assigned with different confidence scores. Several tools were developed and are used within HuBMAP (e.g., Azimuth, Pan-Human Azimuth, FR-Match, RIBCA, STELLAR) and were selected for this reason.

### SC-Transcriptomics CTann Tools

HRApop v1.1 uses five CTann tools (Azimuth, CellTypist, popV, FR-Match, and Pan-Human Azimuth) to assign cell types to sc-transcriptomics data. These tools were chosen as they cover 17 of the 30 organs in the more than 16,000 datasets downloaded from six data portals; were developed within or in support of HuBMAP (Azimuth, Pan-Human Azimuth, FR-Match); and they performed well in a prior algorithm comparison[22]. Plus, these tools come with pre-trained models where the user needs to do minimal refinements. **Fig. 1** shows the number of unique CL terms that can be assigned via these the five CTann tools plus an additional five tools with *skos:exactMatch*. Simple Knowledge Organization System (SKOS) provides standardized mapping properties for expressing relationships between concepts in different vocabularies, ontologies, and terminologies[54]. In this case, *skos:exactMatch* was used when a CTann label was semantically equivalent to an existing CL term and could be used interchangeably. Note that ‘body proper’ (leftmost) refers to all organs in the human body. Pan-human Azimuth covers 440 unique cell type labels belonging to 241 CL terms. Pan-Human Azimuth, DeepCell Types and RIBCA aim to annotate cells in all organs of the human body ('body proper'). Note that FR-Match can only annotate the lung (as of v1.0 of its crosswalk); it was recently extended to also cover retina, brain, breast, gut, skeleton, and fetal immune system (not shown here).

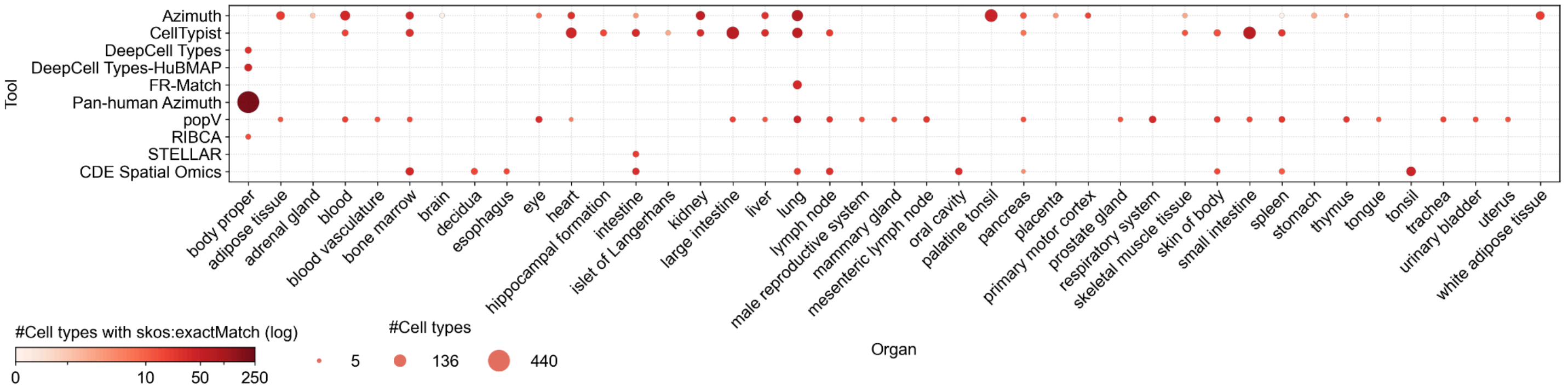

**Figure 1. Number of cell types per CTann and organ combination.** This graph shows the number of cell types that the 10 CTann tools can annotate and the CDE Spatial Omics dataset covers. Note that ‘body proper’ refers to all organs in the human body.

During crosswalk compilation for popV, it was discovered that popV assigns ‘hepatocytes’ to cells in the ‘heart’. However, ‘hepatocytes’ are primary functional cells of the ‘liver’. This issue has also been reported elsewhere[55]. While this issue was present in v1.0 of the popV crosswalk (https://purl.humanatlas.io/ctann/popv/v1.0), it was removed in subsequent releases v1.1-1.4. popV is not used to annotate heart data at all as of HRApop v1.1.

## SC-Proteomics CTann Tools

Three cell annotation tools: DeepCell Types[42], STELLAR[41], and RIBCA[43] have been developed and are used in the HuBMAP and HRA for cell annotation in sc-spatial proteomics datasets. While many cell segmentation tools exist for sc-spatial proteomics data, cell type annotation tools are still at a nascent stage primarily due to resource constraints, e.g., lack of ground truth annotated data which requires much expert time plus funding for computation time to train machine learning models.

## SC-Proteomics CDE Spatial Omics Annotations

Additional single cell spatial omics data was collected in support of constructing the vasculature common coordinate framework. The resulting annotated dataset was named CDE Spatial Omics[56,57] and can be explored in the Cell Distance Explorer (CDE)[12]. CDE Spatial Omics harmonizes cell type data across 14 studies (covering 12 organs) by defining a 3-level cell type hierarchy with input from data providers and cell ontology experts. For the 3-level hierarchy in the CDE Spatial Omics data collection, the most general level (called level 1) contains 8 broad cell types, the next level (level 2) contains 57 unique labels, and the most specific level (level 3) contains 162 unique labels. All cell type labels were mapped to Cell Ontology terms. Different studies in this collection have been segmented and annotated using different methods which are all detailed in their original publications. For the CODEX intestine data, STELLAR was used for cell type annotation. The complete CDE Spatial Omics data collection can be explored on the Spatial Omics Gallery (https://apps.humanatlas.io/cde/gallery).

## Crosswalk Compilation, Validation and Generalization

Ten 'cell type annotation crosswalks' were published in HRA v2.4 (published in December 2025, see https://apps.humanatlas.io/kg-explorer/?do=ctann&versions=v2.4). These 10 crosswalks link CTann tool/author labels to terms and IDs used in CL and are the focus of this paper. Instructions on how to compile cell type (CT) annotation tool crosswalks can be found in the standard operating procedure (SOP) entitled "Authoring Cell Type Annotation Crosswalks"[29]. This section explains the overall process and resulting crosswalks plus the combination of all crosswalks into a "CTann Supertree".

### Identify CL Terms for CTann Labels

As discussed in the introduction, bioinformaticians often use lab- or study-specific cell type labels that deviate from canonical definitions used in ontologies. Examples are 'M2 Macrophage' instead of 'alternatively activated macrophage' or convenience abbreviations like 'T' instead of 'T cell'. Additionally, cell types vary in organ-specificity and generality (depending on the resolution of the technology used). Fully automated approaches to compare cell type annotations are challenging due to these differences in naming conventions and careful human curation by ontology experts is required when comparing and analyzing cell types across labs, organs, and tools.

SKOS was used to create the crosswalks. They provide a set of mapping properties to express relationships between concepts in different vocabularies, ontologies, or terminologies. SKOS mapping properties were used to represent the relationship between CTann tool or human generated cell type annotations and cell type terms in the Cell Ontology (CL): In cases where a cell type could be matched to an existing term, *skos:exactMatch* was used to indicate semantic equivalence and interchangeable use. Novel cell types for which exact matches did not exist were matched to the most appropriate parent term using *skos:narrowMatch* to indicate that the cell type label used by the tool is more specific than the mapped ontology term. For example, cell states are not always included in CL, such as proliferating or activated cells. In the FR-Match tool, 'proliferating transit amplifying cell' was annotated using the CL term, CL:0009010 'transit amplifying cell' as a *skos:narrowMatch*. A total of 531 new cell terms were submitted as new term requests to the CL to provide exact matches in place of the narrow matches. In cases where the tool label is more general than the mapped ontology term *skos:broadMatch* was used to indicate that the cell type label used by the tool is more general than the mapped ontology term (e.g, 'Tuftl' type used by FR-Match tool for lung data, which was more specifically annotated as CL:0002075 'brush cell of trachebronchial tree'). The percentage of exact match, narrow or broad matches in each digital object are described in **Table 2**. We followed best practices for documenting mapping provenance, equivalence versus narrow matching, and uncertainty, similar to the recommendations in the Simple Standard for Sharing Ontological Mappings (SSSOM)[31].

**Table 2: Distribution of SKOS Mapping Relationships (exact, narrow, and broad matches) across CTann Crosswalk Digital Objects in the HRA**

| Tool | Number of Unique Organs | Number of Unique CL IDs | Percent Exact Match | Percent Narrow Match | Percent Broad Match |
|---|---|---|---|---|---|
| Azimuth | 20 | 301 | 57.79% | 42.21% | 0% |
| CDE Spatial Omics | 11 | 126 | 76.8% | 20.2% | 3% |
| CellTypist | 16 | 268 | 57.56% | 42.44% | 0% |
| DeepCell Types | 1 (body proper) | 29 | 100% | 0% | 0% |
| DeepCell Types-HuBMAP | 1 (body proper) | 40 | 97.56% | 2.44% | 0% |
| FR-Match | 1 (lung) | 53 | 59.42% | 37.68% | 2.9% |
| Pan-human Azimuth | 1 (body proper) | 241 | 49.33% | 50.45% | 0.22% |
| popV | 24 | 157 | 99.78% | 0.22% | 0% |
| RIBCA | 1 (body proper) | 15 | 100% | 0% | 0% |
| STELLAR | 1 (intestine) | 21 | 84% | 16% | 0% |

## Crosswalking Process

In June 2026, CL (v2026-06-08) contained 3,322 organ-specific cell terms, providing broad coverage for resolving organ-restricted cell clusters and enabling more precise annotation. Three primary factors guided the crosswalking process:

**Organ Context.** Organ specificity was particularly influential, as many cell annotations were assigned to broad cell type labels that were not organ-specific. When organ metadata were provided by the source dataset, this information was explicitly captured in a dedicated table column (Organ_level) in the table and used to support crosswalking decisions. For example, see the cell type annotations for Azimuth v1.4 in **Table 3** and at https://purl.humanatlas.io/ctann/azimuth/v1.4, where organ terms are mapped to Uberon and include entities such as 'heart' (UBERON:0000948) or 'kidney' (UBERON:0002113).

**Table 3. Crosswalk table format.** Top-20 rows of the Azimuth v1.4 crosswalk. Row 1 shows the title and rows 3-10 with other metadata. Row 11 is reserved for the data header and rows 12-1008 list the data.

| | A | B | C | D | E | F | G |
|---|---|---|---|---|---|---|---|
| 1 | Crosswalk of Azimuth annotated cells to Cell Ontology | | | | | | |
| 2 | | | | | | | |
| 3 | Author Name(s): | Ellen M. Quardokus, Aleix Puig-Barbe, Nicole Vasilevsky, Deanne M. Taylor, Taha Mohseni Ahooyi | | | | | |
| 4 | Author ORCID(s): | 0000-0001-7655-4833, 0000-0001-6677-8489, 0000-0001-5208-3432, 0000-0002-3302-4610 | | | | | |
| 5 | Reviewer(s): | Bruce Herr II, Katy Börner, Jie Zheng | | | | | |
| 6 | Reviewer ORCID(s | 0000-0002-6703-7647, 0000-0002-3321-6137, 0000-0002-2999-0103 | | | | | |
| 7 | General Publicatio | https://azimuth.hubmapconsortium.org/ | | | | | |
| 8 | Data DOI: | https://doi.org/10.48539/HBM469.SKDL.348 | | | | | |
| 9 | Date: | 15-Jun-26 | | | | | |
| 10 | Version number: | v1.4 | | | | | |
| 11 | Organ_Level | Organ_ID | Annotation_Label | Annotation_Label_ID | CL_Label | CL_ID | CL_Match |
| 12 | Heart_L2 | UBERON:0000948 | Adipocyte | AZ:0000001 | adipocyte | CL:0000136 | skos:exactMatch |
| 13 | Heart_L2 | UBERON:0000948 | Arterial Endothelial | AZ:0000002 | endothelial cell of artery | CL:1000413 | skos:exactMatch |
| 14 | Heart_L2 | UBERON:0000948 | Atrial Cardiomyocyte | AZ:0000003 | regular atrial cardiac myocyte | CL:0002129 | skos:exactMatch |
| 15 | Heart_L2 | UBERON:0000948 | B | AZ:0000004 | B cell | CL:0000236 | skos:exactMatch |
| 16 | Heart_L2 | UBERON:0000948 | Capillary Endothelial | AZ:0000005 | capillary endothelial cell | CL:0002144 | skos:exactMatch |
| 17 | Heart_L2 | UBERON:0000948 | Endocardial | AZ:0000006 | endocardial cell | CL:0002350 | skos:exactMatch |
| 18 | Heart_L2 | UBERON:0000948 | Endothelial | AZ:0000007 | endothelial cell | CL:0000115 | skos:exactMatch |
| 19 | Heart_L2 | UBERON:0000948 | Fibroblast | AZ:0000008 | fibroblast | CL:0000057 | skos:exactMatch |
| 20 | Heart_L2 | UBERON:0000948 | ILC | AZ:0000009 | innate lymphoid cell | CL:0001065 | skos:exactMatch |

**Cell typology provided by the tool.** Availability of hierarchical structure in the manual or CTann tool annotations was important for selecting the appropriate level of granularity, particularly in datasets spanning multiple organs (e.g., Pan-Human Azimuth). For selected tools and organ-specific datasets, we also reviewed the original publications and source datasets underlying the annotations to clarify the intended cell types and, where appropriate, identify the corresponding Cell Ontology (CL) terms at the most specific level supported by the source data. In some cases, cells were labeled using intermediate or shorthand terms that lacked explicit anatomical or lineage qualifiers. For example, cells annotated simply as 'neuroendocrine cell' could be crosswalked to the generic term 'neuroendocrine cell' (CL:0000165). When the dataset's annotation hierarchy specified an epithelial lineage within the lung, the cells could instead be named more precisely 'pulmonary neuroendocrine cell' (CL:1000223). Thus, where additional source information was available and needed, the intended cell type described in the underlying publication was considered alongside the tool-provided annotation and its hierarchical context when determining the appropriate CL term. While hierarchical context informs decisions, the current crosswalk tables capture the leaf node (i.e., the most specific term); representation of intermediate annotation levels may be incorporated in future table releases.

**Biomarkers.** Some CTann cell type labels were ambiguous or insufficiently defined for several reasons: (1) an acronym is used that can refer to biologically unrelated cell types (e.g., 'GC' can be interpreted as 'germ cell' (CL:0000586), 'granulosa cell' (CL:0000501) or 'glial cell' (CL:0000125)); (2) reliance on highly specialized or study-specific jargon (e.g., 'branch' for 'migratory enteric neural crest cell' (CL:0002607)); or (3) uncertainty in cell type assignment (e.g., 'doublet like cell'). When contextual information alone was insufficient to resolve these cases, additional curator-driven investigation was required, and marker gene or protein expression profiles proved particularly informative. In instances where ambiguity could not be confidently resolved, cell types were conservatively mapped to a more general parent term to avoid over-specification.

### Manual Review and Exemplification Via Use Cases: Hematopoietic Cells

All cell types underwent manual review with attention to lineage, organ context, and ontological specificity. Here, we use hematopoietic cell types as an illustrative case because their extensive hierarchical structure, numerous closely related subtypes, and variable annotation granularity across resources presented particularly informative challenges for evaluating and refining crosswalk mappings. Hematopoietic cells are immature cells that can develop into diverse types of blood cells, including white blood cells, red blood cells, and platelets. In CL, the term 'hematopoietic cell' includes both immune lineages (leukocytes) and non-immune blood cell types such as 'erythroid' and 'megakaryocytic cells'. Recent multimodal single-cell profiling demonstrates that many immune populations in healthy organs display distinct organ-associated signatures and compositional biases, reflecting adaptations to local microenvironments that are largely conserved across individuals and organs[58]. These organ signatures occur across 'T cells', 'innate lymphoid cells', 'B cells', and 'myeloid lineages', and underscore the limitations of using blood-centric definitions to generalize immune identities in organs.

Where possible, we harmonized immune cell labels to consensus CL terms that reflect both canonical lineage and community usage across datasets of disparate origin. For example, cells annotated in the original datasets as 'CD4 T cells' were mapped to 'CD4-positive, alpha-beta T cell' (CL:0000624), which best aligns with the conventional meaning intended by dataset authors despite not capturing the full functional diversity of CD4 subsets. Similarly, cells labeled as 'NK/T' were mapped to 'mature NK T cell' (CL:0000814), reflecting the mixed invariant T-cell and innate lymphoid biology implicit in those annotations.

Some hematopoietic cells could not be cleanly assigned to a single, well-defined cell type due to mixed lineage signals or insufficient resolution in the original data. In such cases, we identified an appropriate common parent term in the ontology and used *skos:narrowMatch* in the crosswalk to explicitly capture this partial or ambiguous correspondence. For example, cell types annotated as 'erythrocyte/megakaryocyte' (Pan-Human Azimuth dataset) were mapped to 'myeloid cell:erythrocyte/megakaryocyte' (CL:0000763), a composite ontology term that reflects the shared developmental context without asserting equivalence to a more specific cell type. This approach allowed mixed or transitional populations to be represented in a structured and transparent manner while preserving uncertainty and avoiding over-specification.

## Cross-Expert Validation

Cross-validation and agreement analysis were performed using a reproducible, ontology-informed quality-control workflow to harmonize independently curated cell type crosswalk tables. The initial cell type crosswalks were generated by expert curators (1st Curators) and reviewed by team members (1st Reviewers), as reported in **Table 3**. The 1st Reviewers validated the annotations and proposed revisions for mappings where they disagreed with the initial curation. Proposed revisions were reviewed with the curators and incorporated into the crosswalks when consensus was reached. A subset of the resulting crosswalks subsequently underwent blinded re-curation, as described below, to assess inter-annotator agreement and the reproducibility of the curation process.

For each dataset, a pipeline (https://github.com/hubmapconsortium/hra-CTann-crosswalks-validation) ingested the two curator-provided crosswalk tables and validated ontology identifiers and labels against the CL ontology to detect inconsistencies arising from typos, term updates, or misassigned identifiers. Curator-supplied labels were compared to canonical ontology labels and classified as label-correct (exact match), synonym-consistent (match to an ontology synonym), or incorrect, while preserving the original curator annotations. The two tables were then merged (retaining cell types present in either table) and compared to quantify agreement and highlight disagreements. Finally, we used the ontology hierarchy to characterize the ontological nature of disagreements: whether one curator chose a broader term and the other a more specific cell type, or whether they chose different but closely related terms that sit in neighboring branches of the ontology.

**Table 3: Curators and reviewers for each digital object.**

| | 1st round of curation | | 2nd round of curation | | | | |
|---|---|---|---|---|---|---|---|
| **CTann Tool / Dataset** | **1st Curator(s)** | **1st Reviewers** | **2nd Curator(s)** | **2nd Reviewer** | **Initial Inter-annotator agreement** | **Revised to blinded re-curation annotation** | **Original annotation retained** |
| Azimuth | Ellen M. Quardokus, Aleix Puig-Barbe | Bruce Herr II, Jie Zheng | Nicole Vasilevsky | Aleix Puig-Barbe | 65% | 10% | 25% |
| CDE Spatial Omics | Aleix Puig-Barbe | Bruce Herr II | Nicole Vasilevsky | Aleix Puig-Barbe | 88% | 3% | 9% |
| CellTypist | Ellen M. Quardokus | Bruce Herr II, Jie Zheng | Aleix Puig-Barbe, Nicole Vasilevsky | Aleix Puig-Barbe | 69% | 13% | 18% |
| DeepCell Types | Aleix Puig-Barbe | Bruce Herr II, Jie Zheng | Nicole Vasilevsky | Aleix Puig-Barbe | 97% | 0% | 3% |
| DeepCell Types-HuBMAP | Aleix Puig-Barbe | Bruce Herr II, Jie Zheng | Nicole Vasilevsky | Aleix Puig-Barbe | 97% | 0% | 3% |
| FR-Match | Aleix Puig-Barbe | Bruce Herr II, Jie Zheng, Nicole Vasilevsky | N/A | N/A | N/A | N/A | N/A |
| Pan-Human Azimuth | Aleix Puig-Barbe | Bruce Herr II, Jie Zheng | Nicole Vasilevsky | Aleix Puig-Barbe | 69% | 10% | 21% |
| popV | Ellen M. Quardokus | Bruce Herr II, Jie Zheng | Aleix Puig-Barbe, Nicole Vasilevsky | Aleix Puig-Barbe | 94% | 5% | 1% |
| RIBCA | Aleix Puig-Barbe | Bruce Herr II, Jie Zheng | Nicole Vasilevsky | Aleix Puig-Barbe | 94% | 0% | 6% |
| STELLAR | John Hickey | John Hickey, Bruce Herr II, Jie Zheng, Yongqun He, Nicole Vasilevsky | N/A | N/A | N/A | N/A | N/A |

To assess the consistency and reproducibility of the curation process, a subset of crosswalks from each annotation resource underwent blinded re-curation by a second curator. The independently generated annotations were subsequently reviewed by a second reviewer and compared with the original annotations. Inter-annotator agreement for each dataset is summarized in **Table 3**. Following each curation round, the curators met to review annotations and discuss cases of disagreement. Discrepancies were evaluated collaboratively until consensus was reached, resulting in 100% final agreement. **Table 3** reports the initial inter-annotator agreement following blinded curation, as well as the outcomes of the consensus review process, including the proportion of annotations that were revised to the annotation proposed during blinded re-curation and the proportion for which the original annotation was retained. Agreement rates varied across resources, reflecting differences in annotation complexity and granularity. In most cases, disagreements required discussion before consensus was achieved, highlighting the value of independent review in identifying ambiguous mappings and refining crosswalk quality. For example, for the Azimuth crosswalk, initial agreement was 65%. Of the remaining annotations, 10% were revised to the annotation proposed during blinded re-curation, while 25% retained the original annotation.

RIBCA covers only 15 very general cell types which were easy to crosswalk to CL terms. Re-curation surfaced one annotation: 'proliferating/tumor cell' that could be annotated as a *skos:narrowMatch* to ‘neoplastic cell’ or be added as a new term (e.g., ‘proliferating neoplastic cell’). A decision was made to add this new cell type term, see issue in CL (https://github.com/obophenotype/cell-ontology/issues/3581). Until this term is added to CL, a *skos:narrowMatch* to 'neoplastic cell' is used in the RIBCA crosswalk.

The highest inter-annotator variability existed for cases where:

- Very general CT labels are used by the CTann tool, such as schwann (CL:0002573 'Schwann cell' (*skos:exactMatch*)) or Treg (CL:0000815 'regulatory T cell' (*skos:exactMatch*)).
- Very general CT labels are used by the CTann tool and organ-specific CL labels are assigned by the ontology expert, such as 'medullary fibroblast in the kidney' (CL:4030022 'renal medullary fibroblast') or 'astrocytes' in hippocampus CL:0002604 'hippocampal astrocyte' (*skos:exactMatch*)
- Only one biomarker is used to characterize a cell type, for example, 'Alveolar M_ CCL3+' (CL:4033041 'CCL3-positive alveolar macrophage' (*skos:exactMatch*)).

Some recurring reasons for mismatches in inter-annotator agreement included cases such as:

- Annotations were made to a more specific or more general term, such as 'Basal resting' in the lung was annotated to CL:0000646 'basal cell' (*skos:narrowMatch*), which is more general and the preferred annotation was to CL:0002633 'respiratory basal cell:resting' (*skos:narrowMatch*).
- The species-specific (human) annotations were preferred, and at times, a more general (non-species specific annotation) was used and it was agreed to use the more specific cell type. For example, 'Plasmacytoid Dendritic' (CL:0001058 'plasmacytoid dendritic cell, human' (*skos:exactMatch*)).
- The description was vague and we agreed on a different term after revisiting. For example, 'CM Pre-non-Tfh' was originally annotated to CL:0000907 'central memory CD8-positive, alpha-beta T cell:non-T follicular helper' (narrow match) and it was revised to CL:0000904 'central memory CD4-positive, alpha-beta T cell:pre-non-T follicular helper cell' (*skos:narrowMatch*).
- In descriptions where two terms were combined, such as 'Monocyte/cDC', it was agreed upon to annotate up to the higher level concept, such as CL:0011031 'monocyte-derived dendritic cell' (*skos:exactMatch*) or 'Schwann / Neural' was annotated to CL:0002573 'Schwann cell' (*skos:narrowMatch*).
- Cells described as 'like-cells', such as MAIT-like, were difficult to interpret and it was agreed that they should be annotated to the primary cell type, with the 'like' added after the colon. For example, 'MAIT-like' (CL:0000791 'mature alpha-beta T cell:Mucosal-associated invariant T-like cell' (*skos:narrowMatch*)).
- In the original Azimuth dataset, there were some annotations that were to non-CL IDs and these were updated to reflect CL IDs that were added to CL after the original annotations.
- Some of the disagreements were due to simple mistakes, i.e., copy/paste errors by the second curator.

## CTann Supertree in Support of Annotation Comparisons

The 10 crosswalks were used to compile a 'CTann Supertree' that can be used to compare cell type annotations across organs and technologies in support of HRA construction and usage. Level 1 in the CTann Supertree (i.e., the root) is 'cell' just like in the cell type hierarchy for CL (release 2025-12-17)[59], see left in **Fig. 2**. In CL, subclasses of 'cell' include many cell types that are irrelevant for adult human tissue cell type annotation: CL:0000255 'eukaryotic cell', CL:0019001 'tracheobronchial serous cell', CL:0000151 'secretory cell', CL:1000497 'kidney cell', CL:0009010 'transit amplifying cell', CL:4033054 'perivascular cell', CL:0002494 'cardiocyte', CL:0002153 'corneocyte', CL:0002321 'embryonic cell (metazoa)', CL:0000349 'extraembryonic cell', CL:0001061 'abnormal cell', CL:0000163 'endocrine cell', CL:0000064 'ciliated cell' and were excluded; the list of excluded cells is available here:
https://github.com/cns-iu/hra-ctann-supporting-information/blob/main/data/CTann-Supertree-Excluded-Cells.cs

v. Relevant for CTann are 10 cell types from the upper level terms in CL (i.e., the subtypes of the root level ‘cell’) which are the Level 2 CTs in the CTann Supertree:

1. connective tissue cell, CL:0002320
2. endothelial cell, CL:0000115
3. epithelial cell, CL:0000066
4. germ line cell, CL:0000039
5. hematopoietic cell, CL:0000988
6. mesenchymal cell, CL:0008019
7. neural cell, CL:0002319
8. melanocyte, CL:0000148
9. muscle cell, CL:0000187
10. precursor cell, CL:0011115

For the Supertree construction, the most granular cell type from each crosswalk is represented and the lineage is displayed back to the high level superclass from the list above, then the root class, ‘cell’. When cell types in CL had multiple parents, we selected a single path that represented the most granular cell lineage, allowing consistent comparison of annotation granularity across single-cell sequencing tools. To maintain consistency across related cell types, we applied the same lineage selection criteria across cell type groups (e.g., selecting the leukocyte branch for macrophage subtypes). Recursively, all CTs in all 10 CTann crosswalks were added containing 1,468 cell types and a total of 695 unique nodes. The resulting tree was stored in a Google sheet and it is available on GitHub (https://github.com/cns-iu/hra-ctann-supporting-information/blob/main/data/CTann-Supertree-v10.csv).

The CTann Supertree provides a spatial reference framework to overlay and compare the coverage offered by various CTann tools. **Figure 2** shows subtree of the 10 CTann crosswalk Supertree featuring all children nodes of the ‘mature alpha-beta T cell’. Organ specific Azimuth annotations (red nodes) are compared with Pan-Human Azimuth annotations (blue nodes), and cell types identifiable by both Azimuth tools (in purple), reveals distinct patterns: organ-specific Azimuth yields higher-resolution cell type assignments (right side), whereas Pan-Human Azimuth provides broader, more generalized cell type classifications.

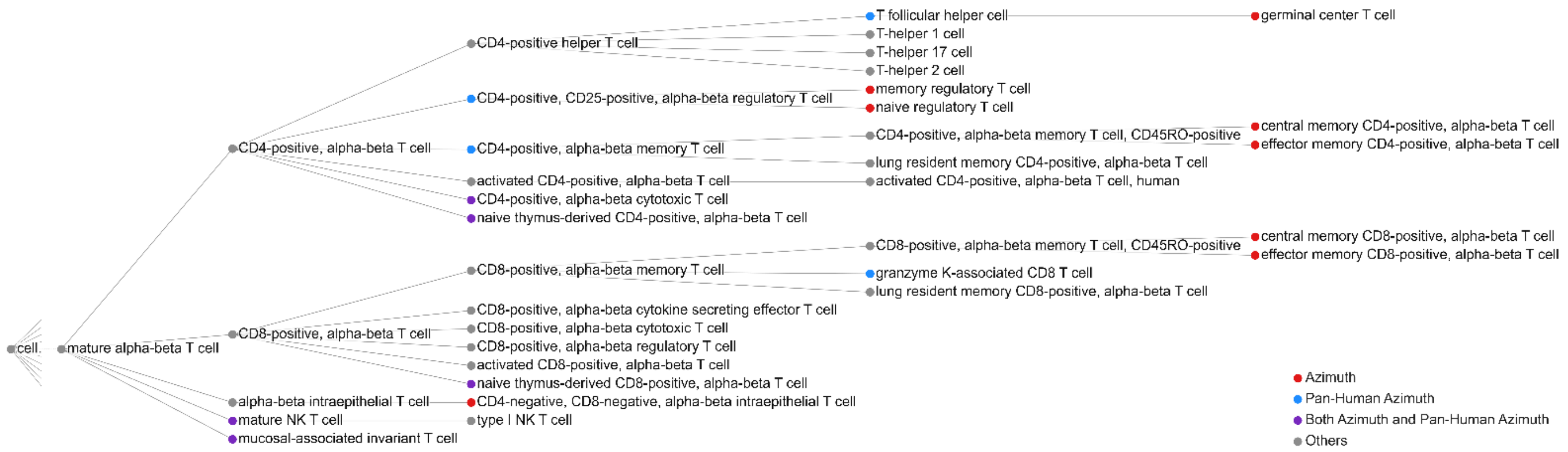

**Figure 2** Partial representation of the CTann Supertree with an overlay of Azimuth (red nodes), Pan-Human Azimuth (blue), and both (purple) annotations to show differences. Explore the high resolution version at https://github.com/cns-iu/hra-ctann-supporting-information/blob/main/code/figures/figure-2-supertree.ipynb.

An **interactive Supertree** visualization is available via the Supplemental Website (https://cns-iu.github.io/hra-ctann-supporting-information/supertree/#overview). The visualization provides four complementary views for exploring and comparing CT labels and their relationships to the CL hierarchy.

1. ***Overview:*** Brief description of what the four different panels show plus data provenance.
2. ***Reference Supertree:*** This view shows how each of the 10 crosswalk tools contributed to the CTann Supertree. The Sources Included table on the left lists the total number of cell types contributed by each tool. In the Supertree, each node is rendered as a fixed-width bar divided into 10 color-coded slots, one per tool, in a constant left-to-right order. A slot is filled with the corresponding tool color when the tool's crosswalk lists that cell type and has no more-specific descendant of it across organs, and remains pale otherwise. 282 of the 695 unique CTs are used by a single tool. 26 general cell types are listed in all 10 crosswalks.
3. ***HRApop Comparison:*** This view shows a comparison of cell types represented in the CTann Supertree with those identified by Azimuth (red) and Pan-Human Azimuth (blue) for lung single-cell transcriptomics data. As depicted in the Direct Tool Outputs in the left sidebar, Azimuth identified 49 distinct cell types and Pan-Human Azimuth identified 95, and the visualization and the Comparison Summary box on the left show there are 12 cell types shared by both tools at the same CL ID. The visualization also captures differences in annotation granularity. For example, CL:0000158 'club cell' is identified by both tools, but Azimuth could identify this node with two different label— 'club (nasal)' and 'club (non-nasal)' – with varying levels of granularity even when the underlying CL ID is the same. Note that this visualization is different from **Fig. 2** that uses all 10 crosswalks.
4. ***HLCA Node Comparison***: Extending the comparison between Azimuth and Pan-Human Azimuth, this tab compares Azimuth and Pan-Human Azimuth cell type annotations with original author annotations in the HLCA dataset. CTann tool annotations are overlaid on the Supertree and filtered by author-defined cell type, allowing non-standardized author labels to be compared with standardized cell type annotations at the individual-cell level. For example, the 4,284 cells labeled by HLCA authors as 'alveolar fibroblasts' are distributed across 30 Supertree cell types predicted by the two methods: 13 unique to Azimuth, 13 unique to Pan-Human Azimuth, and 4 shared. This view highlights how a single author-defined cell type maps across standardized predictions from different tools.
5. ***Tool Agreement:*** This view evaluates agreement among five CTann tools—Azimuth, CellTypist, FR-Match, Pan-Human Azimuth, and popV—using HRApop v1.1 lung sc-transcriptomics data. Because tools may annotate the same population at different levels of specificity, agreement incorporates both exact CL ID matches and descendant relationships: (2 × exact − coverage) / coverage, producing scores from +1 (complete exact agreement) to −1 (no exact agreement). Shape indicates curated annotation difficulty, size indicates tool coverage, and color indicates agreement. All five curated easy cell types score +1.00. Among curated difficult cell types, 'macrophage' and 'pericyte' show the greatest disagreement (−0.60), while 'B cell' scores +0.60. No tool annotates 'goblet cell' or 'secretory epithelial cell'. Across the 53 curated-difficult cell types, mean agreement is +0.80.

## Recommendations for CTann authors

In support of comparisons across organs, technologies, cell type annotation tools or manual methods, we provide the following recommendations:

- **Use the most specific cell type label supported by the data**. Avoid overly general labels when more specific CL terms are available, particularly when organ- or state-specific information is known.
- **Include relevant organ and anatomical context.** Many cell types have organ-specific subclasses in CL and providing anatomical context can improve annotation specificity and consistency.

- **Avoid ambiguous or composite labels.** Labels that combine multiple cell types or biological identities can be difficult to interpret consistently. When possible, select the primary cell type or provide additional context to support a more precise annotation.
    - It should be noted that in some cases, the assay and technology do not provide sufficient evidence to assign a more specific cell type. In particular, fully dissociated tissues lacked spatial context available from approaches using thin tissue sections. Rather than excluding these cells, researchers may choose to retain them at the highest level of specificity supported by the available data.
- **Clearly indicate inferred or similarity-based cell types**. Labels containing terms such as “-like” should clearly identify the primary cell type and the nature of the inferred relationship to reduce ambiguity.
- **Provide sufficient descriptive metadata.** Vague, abbreviated, or underspecified labels were a common source of curator disagreement. Additional information regarding lineage, marker expression, organ context, or cell state can substantially improve annotation quality and reproducibility.
- **Use current Cell Ontology identifiers and terminology.** Periodic review against the latest CL release is recommended, as new terms may become available that better represent the annotated cell type. Using the primary labels or exact synonyms is recommended.
- **Use the CTann Supertree** to compare cell type annotations across tools, organs, and studies. The Supertree supports visualization of annotation coverage and specificity, identification of shared cell types, and comparison of annotations made at different levels of granularity by mapping them to their most specific common ancestor in the Cell Ontology (CL) hierarchy. The interactive visualization provides complementary views for exploring tool coverage, comparing predictions with author annotations, and assessing agreement among CTann tools.

# Data Records

The 10 ‘cell type annotation crosswalks’ are available on Zenodo[60]; the files total 364 KB. They are also available via the HRA KG Explorer: https://apps.humanatlas.io/kg-explorer/?do=ctann. Each crosswalk has a landing page and DOI, e.g., https://doi.org/10.48539/HBM358.CTVQ.639 (CellTypist). All data and SOPs are released under Creative Commons Attribution 4.0 International (CC BY 4.0).

# Technical Validation

To assess the reliability of the cell type mappings, we performed independent blinded re-curation by a second expert curator, in which cell type annotations were independently evaluated by two expert curators, see details in section Cross-Expert Validation in Methods. Curators followed a standardized annotation procedure to map the original cell type labels to the most appropriate Cell Ontology (CL) terms, considering available organ context, supporting publications and biomarker information, and the appropriate semantic relationship (e.g., *skos:exactMatch* or *skos:narrowMatch*). The independently generated mappings were then compared to quantify agreement and identify discrepancies. Discordant annotations were reviewed by the curators and resolved through consensus, providing an expert-based assessment of the reproducibility and robustness of the crosswalk annotations. The SOP further describing this process is available here: https://doi.org/10.5281/zenodo.17873777.

# Usage Notes

The 10 cell type annotation crosswalks are available on the Human Reference Atlas (HRA) Knowledge Graph Explorer website (https://apps.humanatlas.io/kg-explorer/?do=ctann), which lets users search, browse, and download all digital objects in the Human Reference Atlas Knowledge Graph[61]. The crosswalks are available for download in a variety of standard formats that support diverse use cases: tabular (CSV), raw JSON or linked data / semantic web: JSON-LD, Turtle, RDF/XML, N-Triple and N-Quads. These formats enable use in spreadsheets, semantic web tools, graph databases, analysis pipelines, or programmatic workflows. The HRA Knowledge Graph supports SPARQL queries and API access, enabling retrieval of crosswalks and related data dynamically for integration into workflows or analytics notebooks.

Data can be visualized using the ASCT+B Reporter (https://apps.humanatlas.io/asctb-reporter)[1]. The Reporter shows the nested relationships among anatomical structures, cell types, and biomarkers in a tree or indented list view, which allows for exploration of hierarchical relationships and displays where mapped cell types fit within the broader atlas context. Search and filter facets allow for search of specific cell types or annotations within the visualization, which is especially helpful for understanding context or identifying related entities. The ASCT+B Reporter supports comparison of external datasets or crosswalks against reference ASCT+B tables or other datasets, helping validate crosswalks or spot discrepancies. The app provides stats computed from the data and allows exporting those results, enabling integration of visual insights with downstream analysis. The various usages are summarized in **Table 5**.

**Table 4: HRA Crosswalks—Access and exploration**

| Goal | HRA Tool / Format | How to Use |
|---|---|---|
| Browse semantic relationships and detailed metadata | HRA Knowledge Graph Explorer | Search entities, explore links, download in ontology-aware formats |
| Perform structured queries programmatically | SPARQL endpoint / API | Retrieve data in CSV/JSON via scripts, notebooks, or web services[62] |
| Visualize hierarchical anatomy and cell type relationships | ASCT+B Reporter | Use tree views, search, compare across reference tables, export visual outputs |
| Compare CTann across organs, technologies, teams | CTann Supertree Explorer | Select one or more CTann crosswalks to compare coverage and specificity. |

# Data Availability

The following Digital Objects are described in this paper and are available at the link below:

1. Azimuth: https://purl.humanatlas.io/ctann/azimuth/v1.4
2. CellTypist: https://purl.humanatlas.io/ctann/celltypist/v1.3
3. DeepCell Types: https://purl.humanatlas.io/ctann/deepcelltypes/v1.2
4. DeepCell Types-HuBMAP: https://purl.humanatlas.io/ctann/deepcelltypes-hubmap/v1.2
5. FR-Match: https://purl.humanatlas.io/ctann/frmatch/v1.0
6. Pan-Human Azimuth: https://purl.humanatlas.io/ctann/pan-human-azimuth/v1.2
7. popV: https://purl.humanatlas.io/ctann/popv/v1.4
8. RIBCA: https://purl.humanatlas.io/ctann/ribca/v1.0

9. STELLAR: https://purl.humanatlas.io/ctann/stellar/v1.0
10. CDE Spatial Omics: https://purl.humanatlas.io/ctann/vccf/v1.2

The raw Supertee data is available here:
https://github.com/cns-iu/hra-ctann-supporting-information/blob/main/data/CTann-Supertree-v10.csv.

The list of cells excluded from the Supertree is available here:
https://github.com/cns-iu/hra-ctann-supporting-information/blob/main/data/CTann-Supertree-Excluded-Cells.csv.

# Code Availability

The code to generate cross-author validation is available at
https://github.com/hubmapconsortium/hra-CTann-crosswalks-validation.
The code to compute **Fig. 1** and **Fig. 2** is available in the /notebooks dir at
https://github.com/cns-iu/hra-ctann-supporting-information.
The code for the interactive supertree is available in /supertree dir at
https://github.com/cns-iu/hra-ctann-supporting-information.
For reproducibility in HRApop, Docker[63] containers with full contexts for the CTann tools were built, used during HRApop data generation, and made available on GitHub[64]. SPARQL queries to get counts for HRA coverage, such as the number of AS, CTs, and Bs, plus AS modeled in 3D reference organs are available at https://apps.humanatlas.io/api/grlc/hra.html#get-/as-ct-b-counts and https://apps.humanatlas.io/api/grlc/hra.html#get-/as-3d-counts.

# Author contributions

NV led the writing of this paper, reviewed 10 crosswalks a second time, and performed cross-expert validation. APB compiled initial seven crosswalks and reviewed two crosswalks (see crosswalk metadata for detailed author contributions), wrote the pipelines for cross-validation, and performed cross-expert validation. AB leads sc-transcriptomics CTann tool crosswalk usage in support of HRApop releases. EMQ compiled initial three sc-transcriptomics crosswalks. YK visualized two figures. JZ reviewed 9 crosswalks following the initial curation. YH validated the STELLAR crosswalk. DPP implemented the interactive version of the supertree visualization. YJ led sc-proteomics CTann tool crosswalk usage effort for CDE Spatial Omics collection. BWH leads usage of CTann crosswalks in HRA production workflows by reviewing and validating the crosswalks to ensure tool compatibility. KB leads the HRA effort and orchestrated crosswalk creation and usage for mapping new tissue data into the HRA. All authors wrote the paper.

# Competing interests

The authors have no conflicts of interest to declare.

# Acknowledgements

Martin Hemberg provided expert comments on pan-human Azimuth performance. David Osumi-Sutherland provided expertise in CL usage and annotations.

# Funding

The HRA (humanatlas.io) is under active development by HuBMAP, SenNet, the Kidney Precision Medicine Project, GenitoUrinary Developmental Molecular Anatomy Project, and the National Institute of Diabetes and Digestive and Kidney Diseases with expert input by the HRA Editorial Board (humanatlas.io/about#editorial-board) and in close collaboration with experts from 20+ other consortia.

This research has been supported by the following awards:

- HuBMAP:
  - 3OT2OD026671 (APB, AB, EMQ, YJ, BWH, KB)
  - 3OT2OD033756 (NV, APB, AB, EMQ, DPP, JZ, YH, YJ, BWH, KB)
- Whole Person Physiome:
  - U24AT013504 (NV, YK, BWH, KB)
- SenNet:
  - U24CA268108 (AB, YK, DPP, YJ, BWH, KB)
  - U54AG076043 (AB, YJ)
- KMAP:
  - U01DK133090 (BWH, KB)
- Others
  - 1R03OD039970-01 (AB, BWH)
  - CIFAR Catalyst Award (AB, KB)
  - OT2OD030545 (BWH, KB)
  - EMBL-EBI core funds (APB)

The funders had no role in study design, data collection and analysis, decision to publish, or preparation of the manuscript. The content is solely the responsibility of the authors and does not necessarily represent the official views of the NIH.

This research uses IU's computing infrastructure which was supported in part by Lilly Endowment, Inc., through its support for the Indiana University Pervasive Technology Institute.

The Kidney Precision Medicine Project (KPMP) is supported by the National Institute of Diabetes and Digestive and Kidney Diseases (NIDDK) through the following grants: U01DK133081, U01DK133091, U01DK133092, U01DK133093, U01DK133095, U01DK133097, U01DK114866, U01DK114908, U01DK133090, U01DK133113, U01DK133766, U01DK133768, U01DK114907, U01DK114920, U01DK114923, U01DK114933, U24DK114886, UH3DK114926, UH3DK114861, UH3DK114915, and UH3DK114937. We gratefully acknowledge the essential contributions of our patient participants and the support of the American public through their tax dollars.